%% file: paper.tex
\documentclass[10pt,sigconf,letterpaper]{acmart}

\renewcommand\footnotetextcopyrightpermission[1]{} % removes footnote with conference information in first column

\pdfoutput=1

\usepackage{epsfig,endnotes}
\usepackage{color}
\usepackage{caption}
\usepackage{subcaption}
\usepackage{booktabs}
\usepackage{tabularx}   % Should be loaded before arydshln
\usepackage{listings}
\usepackage{arydshln}
\usepackage{array}
\usepackage{makecell}
\usepackage{xspace}
\usepackage{balance}
\usepackage{multirow}
\usepackage{bigstrut}
\usepackage{adjustbox}
\usepackage{wrapfig}
\usepackage[super]{nth} %% % \usepackage[T1]{fontenc}
\usepackage{tikz}
\usepackage{amsmath,amssymb}
\usepackage{stackengine}
\usepackage{enumerate}
\usepackage[utf8]{inputenc}
\usepackage{graphics}
\usepackage{graphicx}
\usepackage{float}
\usepackage{comment}
\usepackage[subtle]{savetrees}
\usepackage{hyperref}

\DeclareUnicodeCharacter{394}{\change}

\hypersetup{
	colorlinks=true,%
	citecolor=black,%
	filecolor=black,%
	linkcolor=black,%
	urlcolor=black,%
}

\newcommand{\mallmixed}{35,645}
\newcommand{\mallmatched}{34,428}

\newcommand{\numreqinit}{160 million}

\newcommand{\numelhits}{\textasciitilde25.51 million}
\newcommand{\numephits}{\textasciitilde50.03 million}

\renewcommand{\shortauthors}{Jueckstock et al.}

\input{includes/commands.tex}

\begin{document}
\title{The Blind Men and the Internet:\\ Multi-Vantage Point Web Measurements}

\author{Jordan Jueckstock}
\affiliation{
	\institution{North Carolina State University}
}
\author{Shaown Sarker}
\affiliation{
	\institution{
		North Carolina State University}
}
\author{Peter Snyder} 
\affiliation{
	\institution{Brave Software}
}
\author{Panagiotis Papadopoulos} 
\affiliation{
	\institution{Brave Software}
}
\author{Matteo Varvello} 
\affiliation{
	\institution{Brave Software}
}

\author{Benjamin Livshits} 
\affiliation{
	\institution{Brave Software, \\ Imperial College London}
}
\author{Alexandros Kapravelos}
\affiliation{
	\institution{North Carolina State University}
}

\input{sections/0_abstract}

\maketitle

\input{sections/1b_alt_intro}

\input{sections/2_motivation}
\input{sections/3_methodology}
\input{sections/4_design}

\input{sections/5_analysis}
\input{sections/6_relatedwork}

\input{sections/7_conclusion}

\balance
\bibliographystyle{plain}
\bibliography{paper}
\end{document}

%% file: includes/commands.tex
\newcommand{\eg}{e.g., }
\newcommand{\ie}{i.e., }

\expandafter\def\expandafter\wrapfigure\expandafter{\wrapfigure{o}{2in}\large\raggedright}

\newcolumntype{L}[1]{>{\raggedright\arraybackslash}p{#1}}

\lstdefinelanguage{JavaScript}{
  keywords={typeof, new, true, false, catch, function, return, null, catch, switch, var, if, in, while, do, else, case, break},
  keywordstyle=\color{blue}\bfseries,
  ndkeywords={class, export, boolean, throw, implements, import, this},
  ndkeywordstyle=\color{darkgray}\bfseries,
  identifierstyle=\color{black},
  sensitive=false,
  comment=[l]{//},
  morecomment=[s]{/*}{*/},
  commentstyle=\color{purple}\ttfamily,
  stringstyle=\color{red}\ttfamily,
  morestring=[b]',
  morestring=[b]"
}

\newif\ifcomment
\commenttrue
\ifcomment
    \newcounter{MVNumberOfComments}
    \stepcounter{MVNumberOfComments}
    \newcommand{\mvnote}[1]{\textcolor{blue}{\small \bf [MV\#\arabic{MVNumberOfComments}\stepcounter{MVNumberOfComments}: #1]}}
    \newcounter{PSNumberOfComments}
    \stepcounter{PSNumberOfComments}
    \newcommand{\psnote}[1]{\textcolor{red}{\small \bf [PS\#\arabic{PSNumberOfComments}\stepcounter{PSNumberOfComments}: #1]}}
\else
    \newcommand\mvnote[1]{}
    \newcommand\psnote[1]{}
\fi

%% file: sections/0_abstract.tex
\begin{abstract}
%Much web privacy, security, and performance research relies on automated measurements of popular websites.
%Researchers typically perform such measurements from well-known vantage points (VPs), e.g., those belonging to universities or cloud services like AWS or Azure.
%While such research is conducted with the assumption (implicit or otherwise) that measurements from these VPs will generalize to the experiences of ``typical’' internet users, there exists both academic and anecdotal evidence that websites respond differently to different client IP addresses.
%Indeed, it is possible that websites (especially illicit or suspicious websites) strategically change their behavior when serving requests from well-known measurement VPs.
In this paper, we design and deploy a synchronized multi-vantage point web measurement study to explore the comparability of web measurements %results 
across vantage points (VPs).
We describe in reproducible detail the system with which we performed synchronized crawls on the Alexa top 5K domains from four distinct network VPs: research university, cloud datacenter, residential network, and Tor gateway proxy.

Apart from the expected poor results from Tor, we observed no shocking disparities across VPs, but we did find significant impact from the residential VP's reliability and performance disadvantages.
We also found subtle but distinct indicators that some third-party content consistently avoided crawls from our cloud VP.
In summary, we infer that cloud VPs do fail to observe some content of interest to security and privacy researchers, who should consider augmenting cloud VPs with alternate VPs for cross-validation.
Our results also imply that the added visibility provided by residential VPs over university VPs is marginal compared to the infrastructure complexity and network fragility they introduce.
\end{abstract}

%% file: sections/1b_alt_intro.tex
\section{Introduction}
\label{sec:introduction}

% explain the literary allusion in our weird title...

An ancient fable from the Indian subcontinent tells of six blind men who set out to study an elephant by feel.
Each encounters a different part of the creature's anatomy (the trunk, a tusk, a leg) and comes away with an absurd generalization of what an elephant is like (a snake, a spear, or a tree).
Taken in isolation, each observation is technically correct but fails to comprehend the whole subject.
While web measurement studies are unlikely to prove as memorable or entertaining, the moral of the story applies.

% Reader's Digest version of original intro's motivation pieces
Researchers regularly measure the state of privacy, performance, and security on the web.
Motivated by cost, convenience, and scalability, most (if not all) researchers conduct large scale measurements of the web from a narrow range of vantage points (VPs), typically cloud systems or research universities.
Prior research and anecdotal experience suggest that certain websites block~\cite{khattak2016you} or cloak themselves~\cite{invernizzi2016cloak} from requests originating from well-known measurement VPs.
Nevertheless, most research proceeds under the tacit assumption that measurements from well known measurement VPs generalize to the web experiences of ``typical'' users browsing from residential networks.
This assumption is a dangerous one; its possible that the kinds of measurements that have been used to motivate improvements to web privacy, security and performance are systematically skewed.
The potential impact to privacy and security is even worse: bad actors may be exploiting this assumption and manipulating the results of the measurements in ways that leave web users vulnerable.

We set out to test the accuracy of this generalizability-assumption by designing a system that can take a range of privacy, performance and security measurements from different web vantage points and compare the results from each VP.  In constructing this multi-vantage point (MVP) study, we take care to document design and implementation decisions so that our system can be  reproduced.
While the web itself is too dynamic to be considered ``reproducible,'' we believe measurement infrastructures and experiment design should be.
We enumerate the anticipated challenges of taking comparable measurements across multiple VPs and justify our design decisions.
We then describe, in detail sufficient to facilitate unambiguous reproduction, the system architecture we deployed and the per-domain and per-page workflow used. 

Deploying our system to crawl the Alexa top 5K domains resulted in \textbf{4TB} of compressed, de-duplicated data collected over 2 weeks.
We present a high-level overview of the experiment results, comparing various metrics for success and failure across crawls and page visits across VPs.
As expected, Tor encountered by far the highest error and failure rates.  But the residential VP proved more fragile than anticipated, with a cascading effect on overall activity volume.
These performance and reliability discrepancies complicate meaningful cross-VP comparisons of metrics such as EasyList and EasyPrivacy hits.
We are, however, able to identify a cluster of structural differences (i.e., changes in 3rd-party content loaded) across VPs that are resilient against performance-mismatch bias and which reveal a modest but tangible blind spot for our cloud VP.

From our preliminary results we infer that security and privacy researchers conducting studies from cloud VPs should consider using additional VP[s] for validation.
Furthermore, while we saw modest but real visibility advantages of our residential and university VPs over the cloud VP, the residential VP provided at best marginal advantages over the university VP given its additional complexity and fragility.

%\mvnote{here it seems that we are generalizing, but we indeed only use one residential VP} \psnote{can we conclude, even hand-wavingly, that findings from university VPs seem to generalize better than cloud VPs?  This would be a meaningful contribution}

In summary, this work makes the following contributions:
\begin{enumerate}

  \item A practical and \textbf{reproducible template} for performing multi-vantage point (MVP) web measurements at scale.
%\mvnote{should we mention here that we will open source as well?}  \psnote{yes!}
  
  \item High-level results and preliminary inferences from a demonstration \textbf{MVP web measurement experiment of the Alexa top 5K} domains, with the raw data to be released on publication.

%\item \psnote{I see that another section has been commented out, but I suggest bringing it back, even if it winds up very short.  Being able to say, even as a prelim-finding, things like ``in our study, university measurements were slightly closer to residential findings than cloud measurements, which can guide implementation decisions in further research`` and ``researchers conducting measurements from residential IPs should take care to account for bandwidth limitations``, etc would be a meaningful contribution.}
  %\item Suggestions and \textbf{guidelines for future work} on how to conduct future web measurements in light of this work.

\end{enumerate}

%% file: sections/2_motivation.tex
%\section{Motivation} \label{sec:motivation}

%% file: sections/3_methodology.tex
%\section{Methodology} \label{sec:methodology}
\section{Design Considerations} \label{subsec:config}

% This section should be a parallel to the next section, each methodology challenge described here should be written in terms of why we need this for the crawl and what actions we should take to resole or address it. The next section drills down to the actual mechanism in the next section.
%\bone{Definition of what constitutes as a vantage point and categorization of vantage points - network endpoints, geo-location, browser profiles etc.}

Web sites can respond to identical requests, from different clients, differently.
% Web-client attributes may directly affect how the web is perceived by such clients. 
Attributes such as originating network, the requesting browser (version and configuration) and operating system (version and configuration), and even hardware characteristics can cause websites to vary their replies. In this paper we compare web measurements taken from four network endpoints (\textbf{vantage points}, or VPs), each with three \textbf{browser configurations}. Our four VPs were: the network of a large research university, a residential ISP endpoint near the university, a cloud endpoint (Amazon EC2) hosted in a regional datacenter about 800 kilometers from the university, and a tunnel provided by the Tor anonymization network. Tor provides an expected worst-case, while the residential and university networks provide reference points against which to compare cloud-based measurements.

We also measured how websites responded given different \textbf{browser configuration} (BC), intended to resemble automated (i.e. crawler) and typical (i.e. human operated) browsers.  All measurements were taken using Chromium 72 running on Linux under the control of Puppeteer~\cite{googlePuppeteer}. The baseline BC is intended to resemble browsers used in most measurement studies by using the Puppeteer defaults, running Chromium in headless mode. The additional BCs used the Xvfb headless display server to run Chromium in full/non-headless mode; one of them also changed the User-Agent string to (falsely) report itself as the same version of Chromium running on Windows 10 in addition to running non-headless.

%\bone{What requirements a multi-vantage point web measurement should have for generalization of the results and how they should be addressed. Why we need to actively minimize the artificial factors to reduce the noise in measurements across vantage points}
%\bone{We want the crawled pages to be chronologically synchronized within a considerable time frame. The contents of the modern web is transient and if not synchronized then can result in significant disparity across vantage points}
%\bone{Also, need to make the web page visit to be as much deterministic across the vantage point as possible, this requires us to have take control over any functionality such as random number generation, we do not control date time functionality as the synchronized page visit should suffice for this.}

As the web changes constantly, differences between two measurements of the same domain may depend on VP, or BC, or the web content itself.
This study aims to understand how changes in VP or BC can affect measurement results; we are not interested in measurement artifacts due to content changes.
To achieve this goal, we take several precautions to minimize measurement-artifacts unrelated to VP or BC.
First, we tear down and restart our automated browser environment for each page visit within each crawl, mitigating variations caused by client-side session state or caches.  Second, We repeat the same crawl-configuration (e.g., Alexa rank, BC, VP) five times to mitigate transient errors.
Third, we take care to measure sites across VP close together in time.
Crawls sharing domain, BC, and repetition count are grouped into \textbf{crawl sets} that are synchronized to launch simultaneously across all four VPs to mitigate time-based content changes.
Fourth, Where our instrumentation relies on randomization, we use a common seed per crawl set.
Finally, to mitigate client-side randomization in web content, we inject JavaScript (JS) logic into each new frame to replace the \texttt{Math}.\texttt{random} API with a deterministic version borrowed from Google's Catapult project~\cite{googleCatapult}.

%\bone{Modern web is very much dynamic, a lot of functionalities are under the hood of the DOM. We need two things for measuring this - a functionality triggering mechanism that somewhat mimics the human behavior on the  page, and a dynamic tracing system that can trace the functionalities getting triggered. The pages the crawler should visit in a nxm crawl, should come from this functionality triggering mechanism.}

%\bone{In a nxm crawl across multiple vantage points, we want each vantage point to crawl the same urls if starting from the same root page. Thus our functionality triggering system should also be identical so that the urls harvested from it across the vantage points are also the same.}

Our unit of work is the \textbf{crawl}, by which we mean a sequence of one or more visited \textbf{pages} within a single domain (eTLD+1).
From each page successfully visited we harvest up to 3 links (the \textbf{width} of the crawl) to identify candidate follow-on pages.
The first page is always the ``landing page" for the domain (\textit{http://DOMAIN.TLD/}) and is assigned a \textbf{depth} of 1.
The crawl stops either when no more queued links are available or until the target depth is reached.
For example, a \textbf{$3 \times 2$} crawl would visit 1 page at depth 1 and 3 pages at depth 2 for a maximum of 4 pages visited per domain.
Crawl dimensions are a design trade-off: more pages means more accurate sampling at the cost of much longer (and more expensive) experiments.
For simplicity, we do not attempt to synchronize the follow-on page URLs visited across crawl sets.

Link harvesting employs seeded random clicking and scrolling activity powered by the \textit{gremlins.js}~\cite{githubGremlins} ``monkey testing'' framework.
This approach allows harvesting of event-triggered navigations with a human-like bias towards large click targets (we do not allow the clicks to actually cause navigation during the page visit).
As a fallback, we also harvest links from HTML anchor tags present in the page document at the end of our visit.
In all cases, links are enqueued only if they navigate to the crawl's original eTLD+1 domain.

%% file: sections/4_design.tex
%\section{Crawling Design \& Data Collection} \label{sec:design}
\section{Crawler Infrastructure}

%\bone{Deployment infrastructure and setup - describe the K8S cluster, database system, queuing mechanism.}

\begin{figure}[t]
    \centering
    \includegraphics[width=0.8\linewidth]{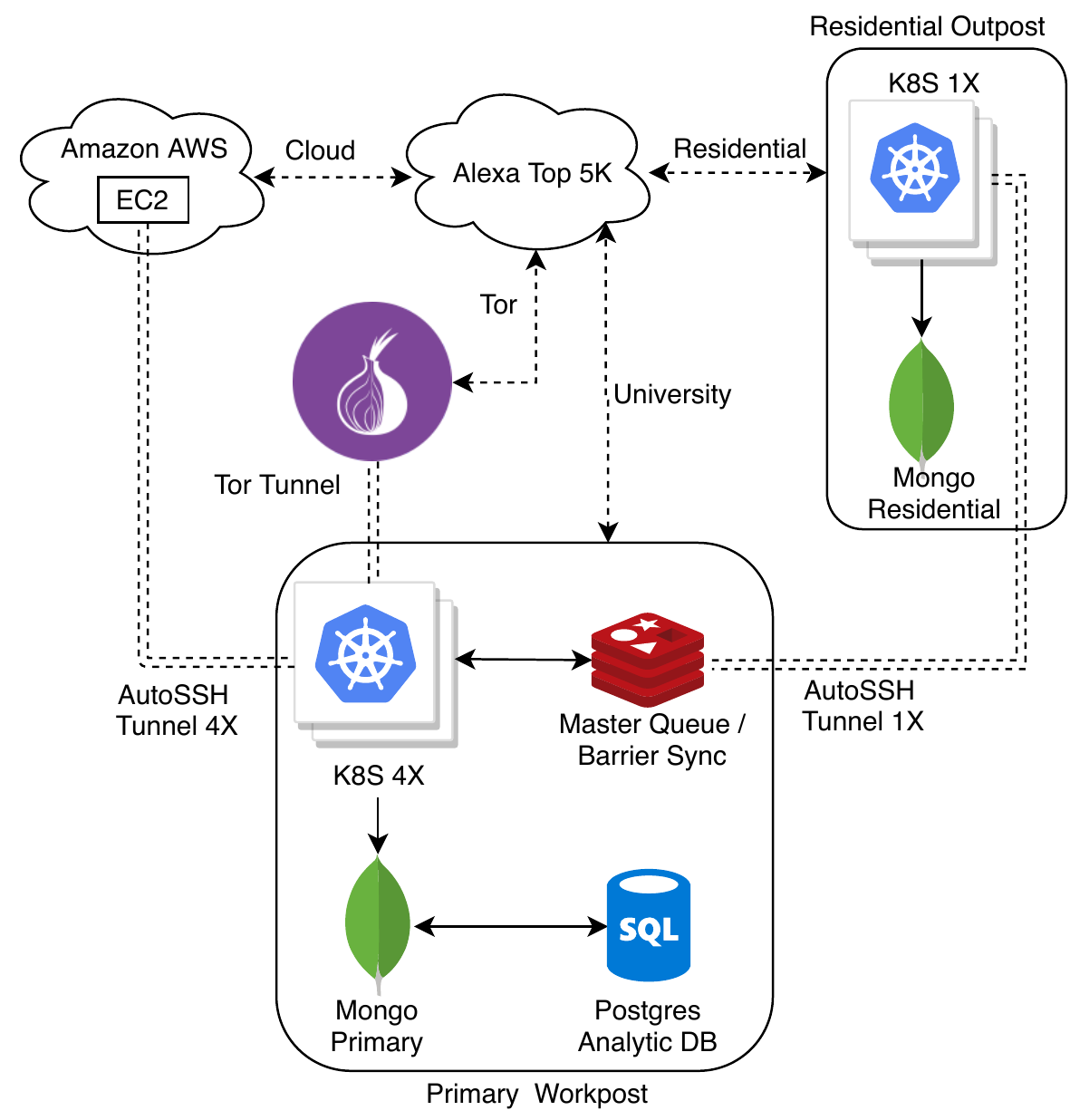}
    \caption{Crawler data collection setup}
    \label{fig:dcoll}
\end{figure}

\noindent\textbf{Architecture}: Our crawling and storage infrastructure was split in two (Figure \ref{fig:dcoll}).
Most of the crawling work and data storage, and all of the centralized control and analysis, were hosted on a primary server cluster hosted within the university network.
This shared cluster---employed by a number of concurrent projects---comprises 8 servers with a total of 144 physical cores and 1 TiB RAM.

Residential crawls and data were hosted on an outpost cluster located in the residence hosting our endpoint.
Resources here were much more modest: crawls performed on a Mac Pro with 8 physical cores and 32GiB of RAM, data stored on a Synology DS918+ NAS appliance.

%\vspace{0.1in}
\noindent\textbf{Crawl Workers}: Crawling was parallelized across many containerized workers running our custom automation framework built on Puppeteer and driving the Chromium browser.
We used a standalone release build of Chromium (72.0.3626.121) rather than the non-release version bundled with Puppeteer.
All data collected or generated during a crawl were stored in MongoDB (one primary server, one outpost server).
Crawl workers were deployed and managed via Kubernetes (K8s). %\mvnote{maybe I missed it but a) what about DNS [added to Vantage Points], 2) was a fresh profile used?} [added to Page Visit]

%\vspace{0.1in}
\noindent\textbf{Crawl Queuing}: Crawl configurations (e.g., domain and browser configuration) were queued to and dispatched from a central Redis server.
A persistent SSH tunnel between the clusters allowed outpost workers to pull jobs from the central queues.
We used a dedicated queue for crawls destined for the residential VP, and a common queue for all other VPs' crawls.

%\bone{Selection of vantage points and browser configurations, and their combinations. Describe each vantage point including how they communicate - over ssh tunnels/sock5 proxies}

%\vspace{0.1in}
\noindent\textbf{Vantage Points}:
University VP crawls required no special configuration: they simply ran in the primary cluster and contacted destinations directly, resolving names via university DNS.
Tor VP crawls required us to configure Chromium to use one of the SOCKS proxies provided by a pool of running Tor clients for communication through the Tor network.
Cloud VP crawls were also tunneled via SOCKS proxy, but this time through a pool of persistent SSH tunnels to a dedicated Amazon EC2 instance.
All tunneled VPs resolved DNS names via SOCKS at the tunnel end-point, using the end-point defaults.

We had originally planned on a tunnel-based approach for the residential VP too, but circumstances intervened: the only readily available residential endpoint featured good downstream capacity (200Mb/s) but was hobbled by low upstream bandwidth (10Mb/s).
So we established the outpost cluster described above and ran residential crawls there directly, resolving names via ISP-provided DNS.
This design change placed a hard limit on our parallelism: the outpost hardware could sustain only 8 concurrent crawls.
This cap effectively limited our global parallelism to 8 workers per VP (32 total) due to crawl set synchronization.

%\bone{How we achieve synchronization across vantage points using the barrier and the description of the barrier architecture}

%\vspace{0.1in}
\noindent\textbf{Crawl Set Synchronization}:
Modern websites change frequently.  To minimize spurious measurement differences between VPs, we start cross-VP crawl sets (same domain, same BC, same repetition) simultaneously.
The synchronization mechanism is a non-resetting barrier implemented using Redis atomic counters and pub/sub notifications.
All workers use the same Redis server as a central broker.
Each crawl configuration includes a \textbf{sync tag} common to all members of a crawl set.
When starting a crawl, workers subscribe to sync notifications for the crawl's sync tag, atomically increment the sync tag's counter, and, if the updated value equals the number of VPs, publishes the ``release'' notification for that sync tag.
All workers in a set, including the release publisher, wait for release notification before proceeding with the crawl.
This mechanism proved robust and effective: all but 11 of the 75,000 crawl sets had all 4 members launch within a \textbf{one second} time window.

\subsection{Collection Workflow}

%\bone{Breakdown into the visiting logic - from page navigation to page load, page loiter and the activities during this - injected scripts for determinism and VV8 tracing, gremlins interactions, link harvesting for next visit during the crawling, and eventual cleanup and/or watch dog timeouts. The crawl matrix size used and the number of repetitions. The possible error cases causing aborts and what counts as a successful crawl and page visit.}

\noindent\textbf{Domain Crawl:} Crawl configurations are pulled from the crawl queues by any available worker.
After synchronization with its crawl set, the worker performs a \textbf{$3 \times 2$} crawl (Section~\ref{subsec:config}) of the target domain using an internal page-visit queue.
Catastrophic errors (e.g., abrupt worker death) result in the crawl ``stalling'' and being re-queued.
Re-queued crawls are detected and marked as \textbf{dropped}.
Crawls are forcibly aborted and the worker process restarted if they do not complete within a watchdog interval of \textbf{180s}.

%%\vspace{0.1in}
\noindent\textbf{Page Visit:} Each page visit starts with launching a fresh browser instance using an empty user profile directory.
Once the new browser is configured and instrumented to capture all relevant data, it is directed to the target URL.
Navigation times out (aborting the visit) at \textbf{30s}; if navigation succeeds, we launch \textit{gremlins.js} pseudo-interaction to collect links.
This interaction runs until we reach 30s from the start of the visit (including navigation time) or until \textbf{10s} have elapsed, whichever is longer.
As we loiter, we close all alert boxes and new windows/tabs opened.
At the end of the interaction window, we perform \textbf{tear-down}: capturing the final state of the DOM, taking a screenshot, extracting anchor tag links from the DOM, and saving any other data we have been collecting in memory during the visit.
Tear-down must complete within \textbf{5s} or the visit is aborted.
The worst-case duration, then, of a single page visit is \textbf{45s}.
All timeouts and limits were set pragmatically, derived from the desired volume of our experiment and the constraints of our time budget, and are comparable to counterparts in prior measurement work~\cite{snyder2016browser}.

%\bone{Broad overview of collected data - all events fired during page time information and the relevant data collected, information collected on page, frame, script, request and the corresponding responses.}
%\vspace{0.1in}
\noindent\textbf{Data Collected:} In addition to crawl and page visit metadata (configuration details, lifecycle timestamps, final status), we record web content data and metadata for each page visited, including: the HTML frame tree (parents, children, and navigation events), all HTTP requests queued within Chromium (URL, resource type, frame context), HTTP request responses (status, headers, and body) and failures (error type).

%\vspace{0.1in}
\noindent\textbf{Post-processing:} After collection was complete, we exported raw collection data from the MongoDB servers into a PostgreSQL server for data aggregation and relational analytics.
We further post-processed the HTTP request metadata (document URL, request URL, and resource type), matching against the popular EasyList (EL)~\cite{abpEasyList} and EasyPrivacy (EP)~\cite{abpEasyPrivacy} AdBlockPlus (ABP) filter lists.
Finally, we post-processed each page's captured DOM content and related request data to search for indications of CAPTCHA presence using rules extracted from Wappalyzer~\cite{githubWappalyzer}.

\begin{comment}
\begin{figure}
    \centering
    \includegraphics[width=0.9\linewidth]{figs/crawl_lifespan.pdf}
    \caption{Events in a crawling job lifespan}
    \label{fig:jspan}
\end{figure}
\end{comment}

%\bone{Summary numbers for collected data - no of crawls succeeded/aborted, pages visited, requests, frames, scripts, events, and the blobs if necessary. The timeline when data was collected and which Alexa slice crawled. This are raw numbers here only, we dig deep into these in later sections}

%\vspace{0.1in}
\noindent\textbf{Collection Scope \& Timeline:} We crawled the top \textbf{5,000} Alexa domains as reported on 2019-04-01~\cite{alexaTopMillion}. Each domain was crawled \textbf{5} times using \textbf{3} BCs across \textbf{4} VPs, for a total of \textbf{300,000} crawls, each visiting a maximum of \textbf{4} pages (for a maximum of \textbf{1,200,000} page visits).
Crawling commenced 2019-04-07 and finished 2019-04-20.
The EL and EP lists used in post-processing date from 2019-04-18.

%% file: sections/5_analysis.tex
\section{Analysis \& Results} \label{sec:analysis}
\begin{figure}[t]
    \centering
    \includegraphics[width=0.9\linewidth]{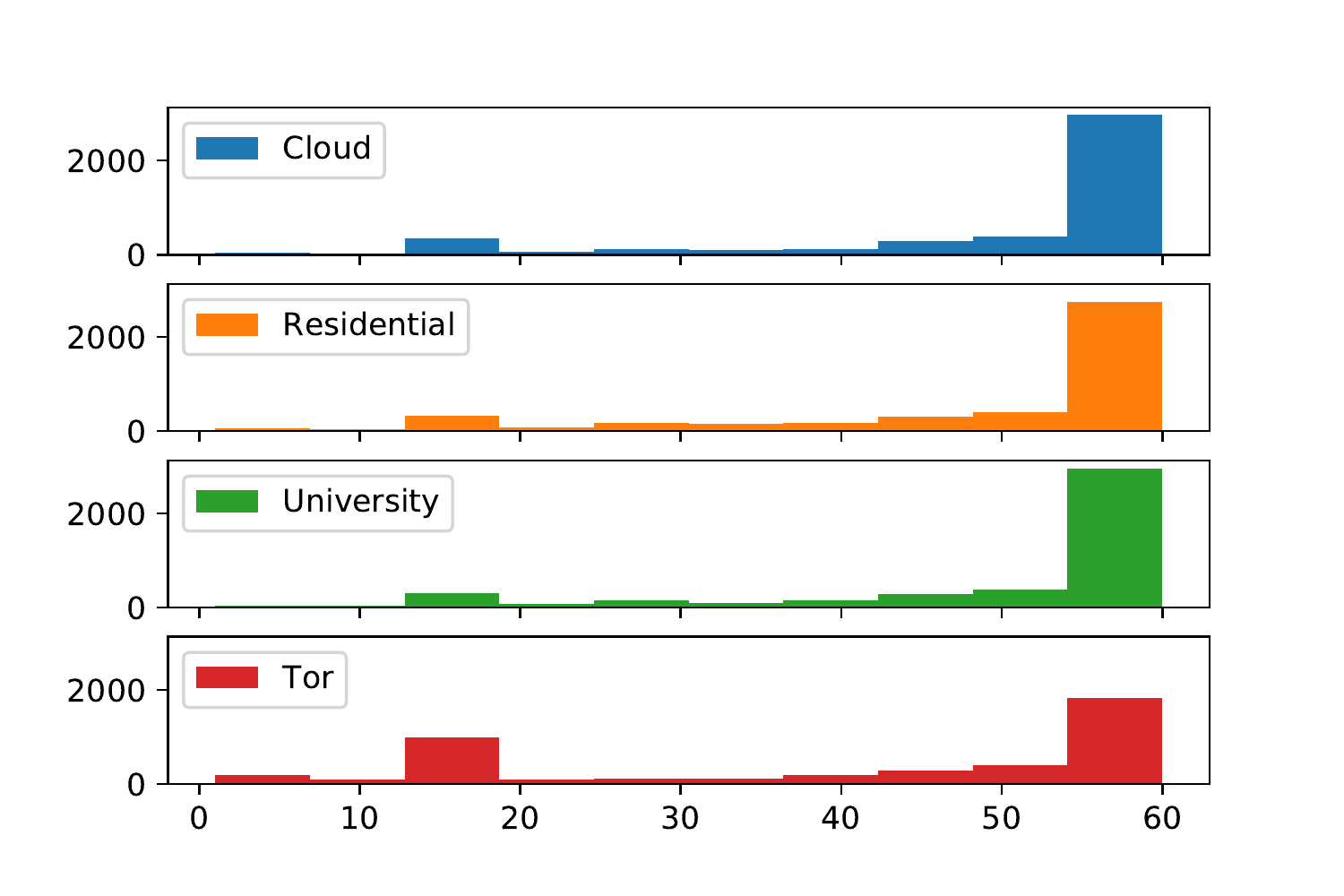}
    \caption{Completed pages with no dead-end across vantage points}
    \label{fig:hist-no-de}
\end{figure}

\noindent\textbf{Summary of Collected Data:} During our crawl, for each of the Alexa top 5K domains, we visit 4 URLs including the root page URL. Each crawl is of depth 2 and width of 3 where the root page is depth 0 and we find the 3 URLs to visit for the next depth from the root page. As each URL was visited 5 times over 3 browser configurations, it resulted in a maximum of 60 page visits for each crawl on a single VP given all page visits were successful. We define a page visit to be a dead-end if the visit finished with no HTML page content. In Figure~\ref{fig:hist-no-de}, we show the distribution of the page visits for the Alexa domains excluding the dead-end pages over all VPs. All VPs show a bimodal distribution - one of the modes being at the 60 pages mark which is expected, the other at 15. Since we visit each page 5 times over 3 browser configurations the mode at 15 indicate no other pages beyond the root page was visited during the crawl. This is particularly noticeable in case of Tor, where the likelihood of having a captcha page returned for the root page is higher compared to other VPs (approximately 12\% of all Tor pages had captcha presence compared to \textasciitilde8\% on other VPs), with no URL harvesting possible.

% Binning of crawls by successful page visits 
We consider a page visit during crawling to be completed when the visit succeeded without any error and the page was marked finished. Based on this, we categorized our crawls by the number of completed page visits across VPs into three classes over the browser configuration used: \textit{none} crawls are those for which none of the page visits were completed, \textit{mixed} and \textit{matched} are those for which the number of page visits are different and exactly the same across VPs respectively. Table~\ref{tbl:bin-crawl} shows the breakdown of crawls excluding Tor. When Tor is included, the number of \textit{mixed} crawls increases to \mallmixed~and \textit{matched} decreases to \mallmatched~due to Tor's overall inconsistency of page visits during crawls.  Crawl set mix and match ratios provide one of the only high-level results that show clear if minor distinctions between browser configurations (BCs): the headless BC resulted in more matched sets compared to the more realistic but more heavy-weight ``headed'' BCs.

\begin{table}[t]
    \begin{tabular}{l|r|r|r}
    \hline
    \begin{tabular}[c]{@{}l@{}}Crawls by\\ Page Visited\end{tabular} & \multicolumn{1}{l|}{\begin{tabular}[c]{@{}l@{}}Headless \\ Linux\end{tabular}} & \multicolumn{1}{l|}{\begin{tabular}[c]{@{}l@{}}xvfb with\\ UA:Windows\end{tabular}} & \multicolumn{1}{l}{\begin{tabular}[c]{@{}l@{}}xvfb with\\ UA:Linux\end{tabular}} \\ \hline
    None                                                                & 1,779                                                                           & 1,870                                                                                & 1,855                                                                             \\ 
    Mixed                                                               & 5,310                                                                           & 6,547                                                                                & 6,519                                                                             \\ 
    Matched                                                             & 17,911                                                                          & 16,583                                                                               & 16,626                                                                            \\ \hline
    \textbf{Grand Total}                                                & \textbf{25,000}                                                                 & \textbf{25,000}                                                                      & \textbf{25,000}                                                                   \\ \hline
    \end{tabular}
\caption{Crawls over configurations categorized by number of pages visited successfully across vantage points excluding Tor}
\label{tbl:bin-crawl}
\end{table}

% Breakdown of aborts
In Table~\ref{tbl:cause-abort}, we display the major causes of page abort by percentage across VPs. The page navigation timeout was the most prominent cause of page aborts during crawls. The absence of DNS name resolution failure over the cloud and Tor, and symmetrically the absence of socks proxy connection error over residential and university indicates that these errors are from the same class. Due to cloud and Tor being connected to the primary cluster over a tunnel the DNS errors were perceived as SOCKS proxy errors over these VPs. 

\begin{table}[t]
\resizebox{\linewidth}{!}
{
    \begin{tabular}{l|r|r|r|r}
    \hline
    Error Type                                                                   & \multicolumn{1}{l|}{Cloud} & \multicolumn{1}{l|}{Residential} & \multicolumn{1}{l|}{Tor} & \multicolumn{1}{l}{University} \\ \hline
    \begin{tabular}[c]{@{}l@{}}DNS Name \\ resolve failed\end{tabular}           & N/A                      & 4.26\%                           & N/A                    & 3.26\%                         \\ \hline
    \begin{tabular}[c]{@{}l@{}}SOCKS proxy \\ connection error\end{tabular}      & 3.84\%                     & N/A                              & 4.05\%                   & N/A                            \\ \hline
    \begin{tabular}[c]{@{}l@{}}Timeout during \\ post-visit cleanup\end{tabular} & 3.50\%                     & 7.42\%                           & 2.98\%                   & 3.42\%                         \\ \hline
    \begin{tabular}[c]{@{}l@{}}Timeout exceeded\\ for navigation\end{tabular}    & 9.32\%                     & 13.36\%                          & 19.90\%                  & 10.69\%                        \\ \hline
    Other                                                                        & 3.18\%                     & 3.72\%                           & 3.12\%                   & 3.99\%                         \\ \hline
    \textbf{\% of total failures}                                                & \textbf{19.84\%}           & \textbf{28.75\%}                 & \textbf{30.05\%}         & \textbf{21.36\%}               \\ \hline
    \end{tabular}
}
\caption{Breakdown of abort causes of crawls}
\label{tbl:cause-abort}
\end{table}

\begin{figure}[t]
    \centering
    \includegraphics[width=0.85\linewidth]{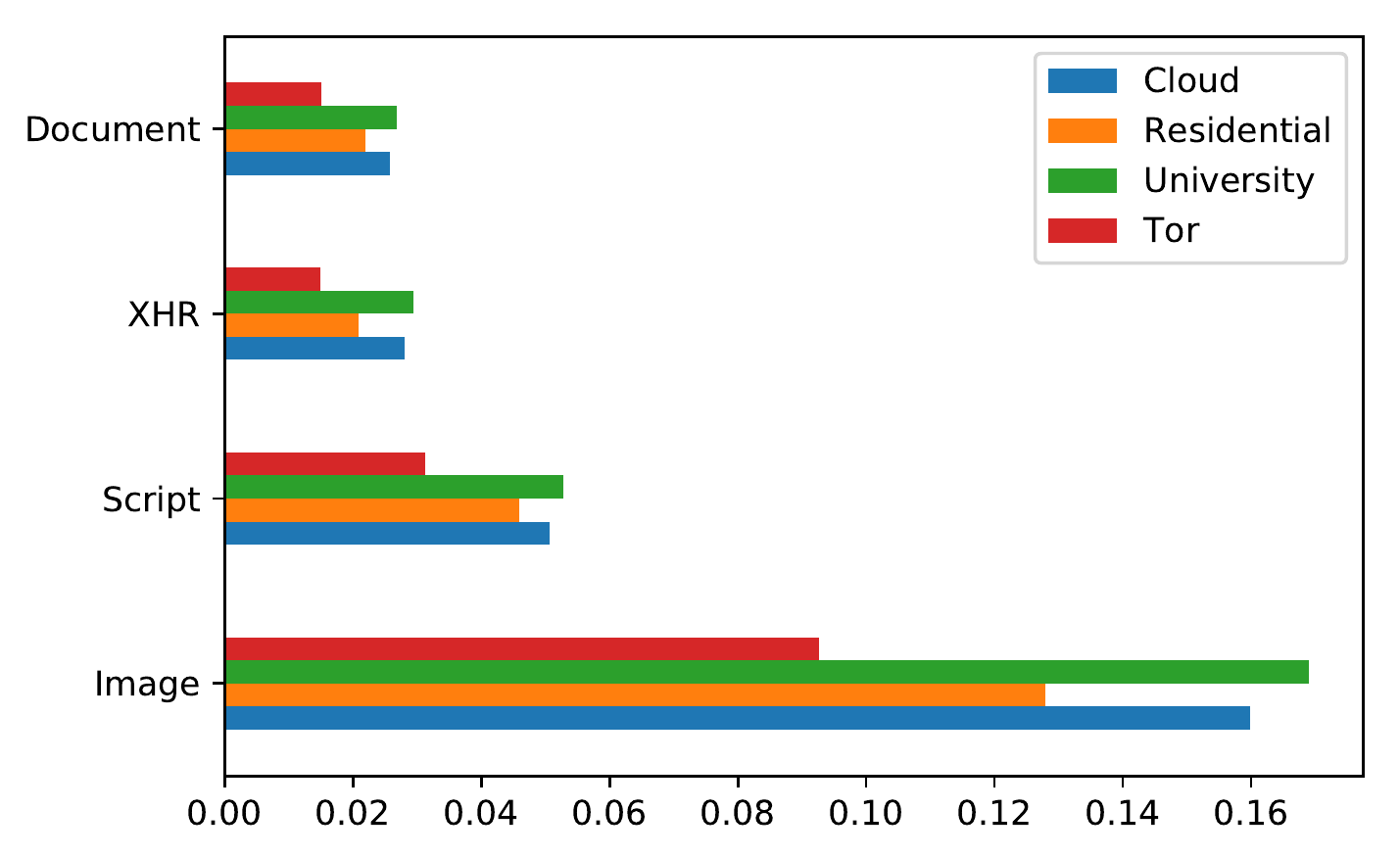}
    \caption{Initialized requests, percentage by resource type across vantage points}
    \label{fig:req-init-vp-rt}
\end{figure}

During our data collection, we collected approximately~\numreqinit\xspace request initializations over all VPs. Figure~\ref{fig:req-init-vp-rt} displays the percentage breakdown of the top 5 request initializations by their requested resource type across the VPs. Images were the largest request type covering more than 50\% of the entire collected request initializations. Among VPs, excluding Tor, residential had lower number of requests compared to the other two - cloud and university, which is evident also in the figure for resource types image, script, and XHR. This can be partially explained by the lack of network bandwidth for the residential VP. To understand this empirical observation further, we looked into the request initializations filtered by the EasyList and EasyPrivacy URL filters. 

\begin{figure}[t]
    \centering
    \includegraphics[width=0.85\linewidth]{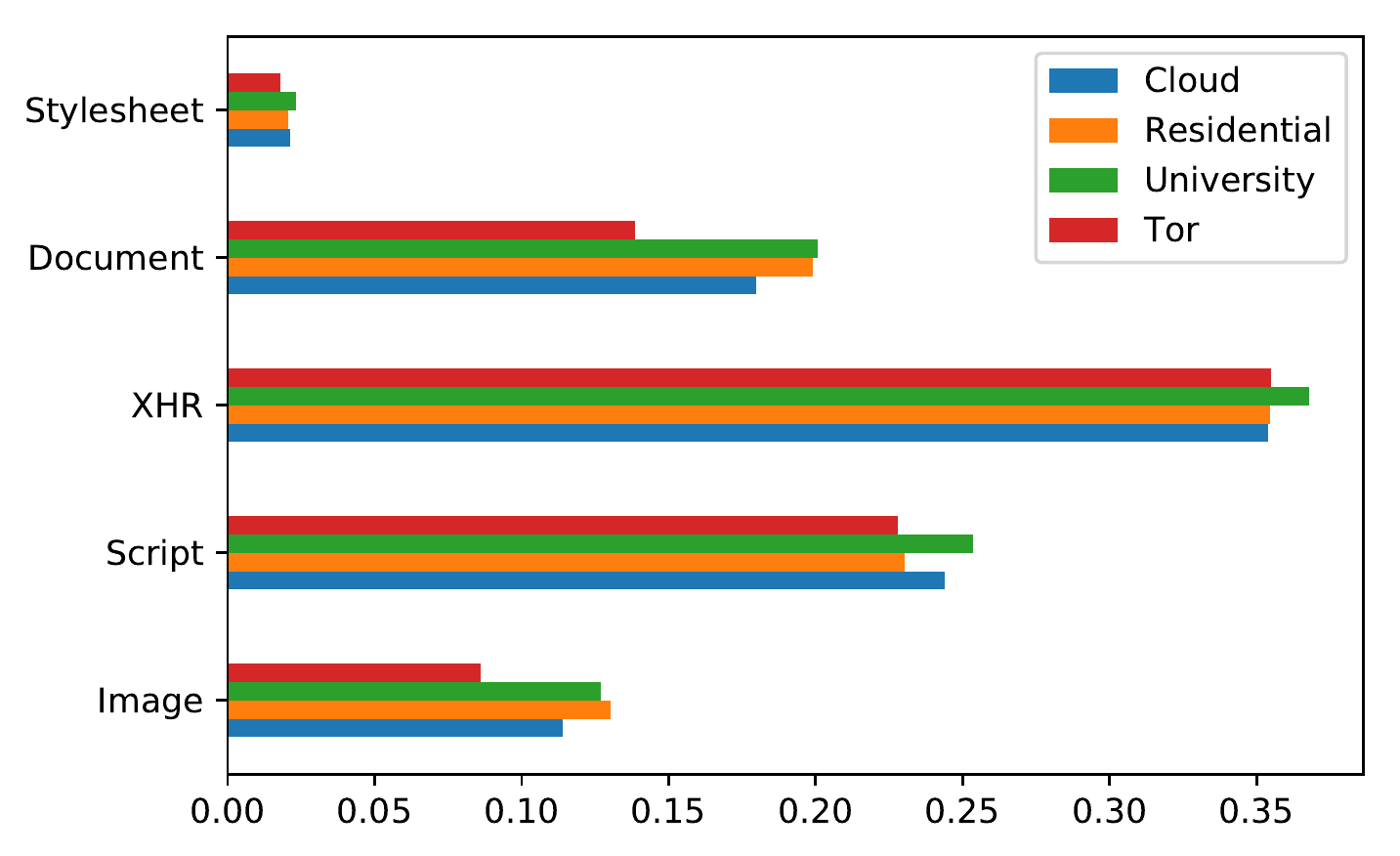}
    \caption{Ratio of EasyList requests to total requests across vantage points by resource types}
    \label{fig:ratio-el-vp}
\end{figure}

\noindent\textbf{URL Hits on Filter Lists:} We matched \numelhits\xspace and \numephits\xspace requests against EasyList (EL) and EasyPrivacy (EP) respectively. Figure~\ref{fig:ratio-el-vp} depicts the ratio of the number of request initializations that matched the rules on the EL filter to the total request initializations across VPs, categorized by the prominent resource types. From the figure, we can see that despite the residential VP being limited in terms of network bandwidth and capacity, it had similar (more in case of images) ratio of matches for document and image resource types. As HTML content (document resources) and images form the majority of ads on the web and EL filter being an ad blacklist, this implies cloud receiving lesser ad content compared to residential and university VPs.

However, we observe differently on Figure~\ref{fig:ratio-el-vp}, which displays the similar ratio between EP matches over all VPs, divided by the top resource types. Here we see a decrease in image type resources being matched for residential compared to other vantage points. EP filter consists of tracker blacklisting rules, and combining with the common practice of web trackers to leverage of tiny images for ad tracking~\cite{googlpxtrk, fbpxtrk}, this drop in image resource matches on EP filter points towards increased presence of trackers over VPs other than residential.

The disparity of these results shows the pitfall of volumetric analysis of requests and at the same time indicates the presence of discrimination on VPs from the web. A better approach to explore and understand this discrimination would be to examine the subset of third-party requests across vantage points, as both the EL and EP filters are comprised of mostly third-party resource rules.

\begin{figure}[t]
    \centering
    \includegraphics[width=0.85\linewidth]{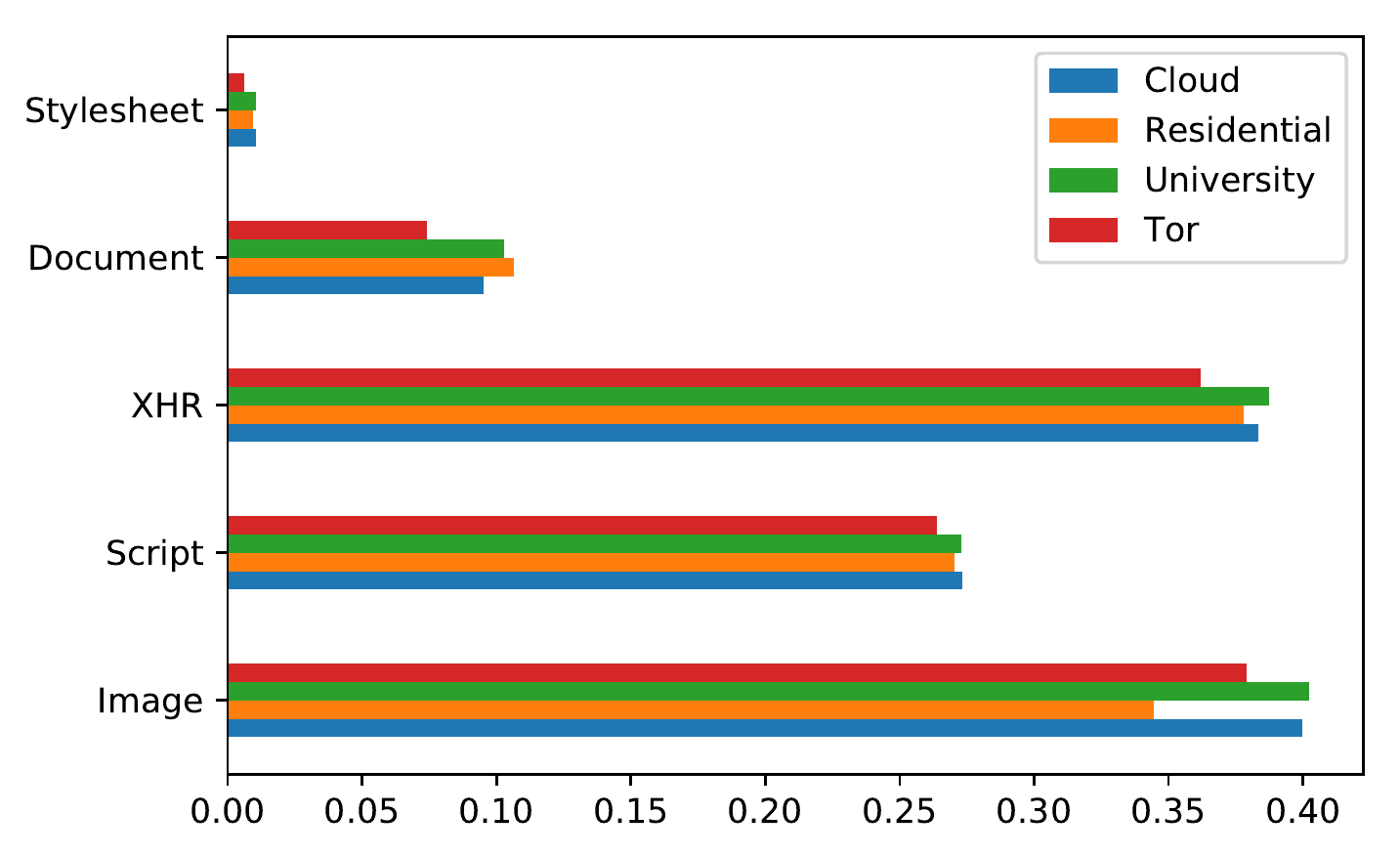}
    \caption{Ratio of EasyPrivacy requests to total requests across vantage points by resource types}
    \label{fig:ratio-ep-vp}
\end{figure}

\begin{figure}[t]
    \centering
    \includegraphics[trim=0.5cm 1.3cm 0.5cm 1.6cm, clip, width=0.9\linewidth]{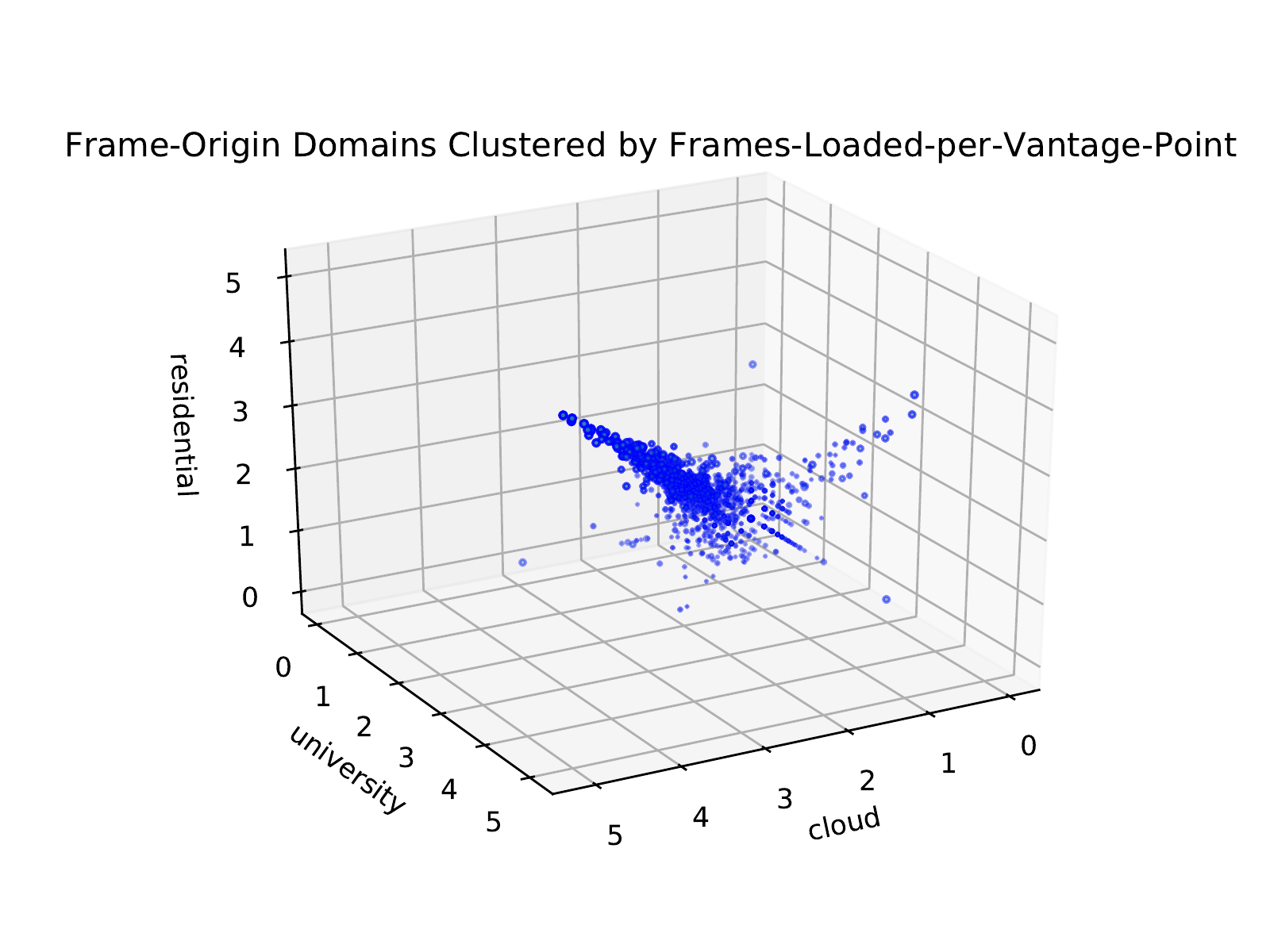}
    \caption{Visual clusters of VP-exclusive frame origin domains}
    \label{fig:frame_origin_clusters}
\end{figure}

\noindent\textbf{Comparison of Vantage Points:} Figure \ref{fig:frame_origin_clusters} provides a microcosm of our results.
Each marker denotes a distinct origin domain from which we observed third-party sub-frames loaded. The three axes ($\log_{10}$) denote volume of 3rd-party frames loaded for visitors from a given VP (as only 3 dimensions are available, Tor is excluded as the least interesting VP for general-purpose measurement).
The prominent, tapering spike of markers extending toward the top left rides the \textbf{central diagonal} of the plot: it comprises origins frequently loaded for visitors from all VPs.
A few outliers and strings of markers lie in the lower \textbf{corners} of the plot: origins loaded exclusively for visitors from a single VP.
The only other visually significant structure of markers is a faint but unmistakable spike of markers climbing the cloud-0 wall: origins that loaded somewhat frequently for visitors from residential and university VPs, but \textbf{never} for cloud visitors.
It is unlikely that blind spot is simply a side-effect of localized ad targeting, since there is no cluster of comparable range/mass along the cloud-only axis.

%% file: sections/6_relatedwork.tex
\section{Related Work} \label{sec:relwork}
%This section describes prior work on measuring how websites discriminate, or
%vary their content, depending on characteristics of the visitor.  We
%intentionally \textit{do not} go into work focusing on censorship, as that work
%focuses on a similar but distinct problem.  The censorship work (broadly)
%focuses on instances of governments or network operators preventing users from
%visiting certain types of content, independent of the wishes of the content
%host.  This work, and the disused related work, all focus on the inverse
%problem; situations where the content host hides or manipulates its content,
%depending on the visitor.

%MV -- found it verbose, so compressed a bit 
This section describes prior work on measuring how websites discriminate, or
vary their content, depending on characteristics of the visitor.  We
intentionally \textit{do not} discuss work focusing on censorship, as they focus on the inverse problem, \ie governments or network operators preventing users from visiting certain types of content, independent of the wishes of the content
host.  %This work, and the disused related work, all focus on the inverse problem; situations where the content host hides or manipulates its content, depending on the visitor.

%\subsection{Distributed measurement systems}
%\vspace{0.1in}
\noindent\textbf{Distributed Measurement Systems:} Much prior work has focused on the design and deployment of systems for
detecting when networks and site providers discriminate based on visitor
attributes, primarily IP address.  Bajpai et al.~\cite{bajpai2015survey} provide
a summary of this work, including the strengths, differences, and lineages of
existing proposals. In our study, we are concerned about measuring how the web reacts when visited from different VP endpoints.

PacketLab~\cite{levchenko2017packetlab} proposes a universal measurement endpoint system by decoupling the measurement logic from the actual system and adopting an access control system for the physical endpoints. In contrast, our architecture is not concerned about endpoint network infrastructure as a packet source/sink, but distributed VPs measuring web content.

%\subsection{Website discrimination}
%\vspace{0.1in}
\noindent\textbf{Website Discrimination:} Other related work focuses on understanding the motivations behind, and frequency of, websites presenting different content to different users.
Some researchers have focused on understanding when, why and how websites
block IP addresses for security reasons.  Khattak et al.~\cite{khattak2016you}
explores how websites treat requests coming from the Tor network differently
than ``standard'' internet traffic.  The work visits the 1k most popular
websites and compares how websites respond differently to Tor and non-Tor
requests.  This work is similar to the Tor-related measurements in our work,
though over a smaller number of websites (1k versus 5k).  Afroz et al.~\cite{afroz2018exploring}
found that a significant amount of IP-based blacklisting is likely unintended,
and the result of overly-general security policies on networks.
Tschantz et al.~\cite{tschantz2018bestiary} looked into a variety of motivations for IP based blocking and found that security was a major motivation,
along with political (\ie GDPR) reasons.

Invernizzi et al.~\cite{invernizzi2016cloak} investigated security-motivated IP
based discrimination from the inverse security motivation; websites attempting
to hide their malicious activities (instead of websites shielding themselves
from malicious visitors).  The authors found a large number of websites using
IP lists to show benign content to visitors coming from well known measurement
IPs, while showing malicious content to other (assumed to be human) traffic.

Additional research has explored the motivations for websites
presenting different content to users based on their IP addresses.% based on the visitors IP address.
Fruchter et al.~\cite{fruchter2015variations} found that websites track users
differently, and to varying degrees, based on the regulations of the country
the visitor's IP is based in.  Iordanou et al.~\cite{iordanou2017fiddling}
describes a system for measuring n how e-commerce websites discriminate between
users.  The authors consider several different motivations for discrimination,
including geography (measured by IP address), prior browsing behavior (\eg
tracking-derived PPI) of the user, and site A/B testing.  The authors find that
the first and third motivations explain more site ``discrimination'' than the
second motivation.

%% file: sections/7_conclusion.tex
\section{Conclusion} \label{sec:conclusion}

We conclude with a few recommendations for future MVP measurement studies.
Such studies should implement the controls we enumerate in Section~\ref{subsec:config} to minimize false differences across VPs.
Additional controls to consider include throttling to achieve bandwidth and compute resource parity across remote VPs.
Given limited choice in class of VP, prefer research university networks to cloud systems, as the former appear to generalize slightly better to true residential browsing experiences.
Security and privacy researchers in particular should beware of possible blind spots in third-party content coverage from cloud VPs.
We look forward to releasing our collected data to the measurement research community.